\documentclass[aps,preprint,superscriptaddress,nofootinbib]{revtex4-1}
\usepackage{epsfig,amsmath,amssymb,verbatim,mathrsfs,array,layout,textcomp,amssymb,latexsym,color,multirow}

\newenvironment{Eqnarray}{\arraycolsep 0.14em\begin{eqnarray}}{\end{eqnarray}}
\def\beqa{\begin{Eqnarray}}
\def\eeqa{\end{Eqnarray}}
\def\beq{\begin{Eqnarray}}
\def\eeq{\end{Eqnarray}}

\newcommand{\no}{\nonumber}
\newcommand{\bv}{\left(\begin{array}{c}}
\newcommand{\ev}{\end{array}\right)}
\newcommand{\bmtwo}{\left(\begin{array}{cc}}
\newcommand{\bmthree}{\left(\begin{array}{ccc}}
\newcommand{\emn}{\end{array}\right)}
\newcommand{\bmtwoc}{\left\{\begin{array}{cc}}
\newcommand{\bmthreec}{\left\{\begin{array}{ccc}}
\newcommand{\emnc}{\end{array}\right\}}
\newcommand{\ba}{\begin{array}}
\newcommand{\ea}{\end{array}}

\newcommand{\Ghg}[2]{\Gamma_{(#1)_{\perp}, #2 }}

\newcommand{\abs}[1]{\left\lvert#1\right\rvert}
\definecolor{red1}{cmyk}{0,1,1,0.3}

\def\lsim{\mathrel{\rlap{\lower4pt\hbox{\hskip1pt$\sim$}}
     \raise1pt\hbox{$<$}}}         
\def\gsim{\mathrel{\rlap{\lower4pt\hbox{\hskip1pt$\sim$}}
     \raise1pt\hbox{$>$}}}         

\def\babar{\mbox{\slshape B\kern-0.1em{\small A}\kern-0.1em B\kern-0.1em{\small A\kern-0.2em R} }}

\begin{document}

\title{
Subtleties in the \babar measurement of time-reversal violation\vspace*{4mm}}

\author{Elaad Applebaum}
\email{eappleb2@illinois.edu}
\affiliation{Department of Physics,
University of Illinois at Urbana-Champaign, Urbana, Illinois 61801, USA\vspace*{2mm} }

\author{Aielet Efrati}
\email{aielet.efrati@weizmann.ac.il  }
\affiliation{Department of Particle Physics and Astrophysics
Weizmann Institute of Science, Rehovot 7610001, Israel\vspace*{2mm}}

\author{Yuval Grossman}
\email{yg73@cornell.edu}
\affiliation{Laboratory for Elementary-Particle Physics, Cornell University, Ithaca, New York 14853, USA\vspace*{2mm}}

\author{Yosef Nir}
\email{yosef.nir@weizmann.ac.il  }
\affiliation{Department of Particle Physics and Astrophysics
Weizmann Institute of Science, Rehovot 7610001, Israel\vspace*{2mm}}

\author{Yotam Soreq}
\email{yotam.soreq@weizmann.ac.il  }
\affiliation{Department of Particle Physics and Astrophysics
Weizmann Institute of Science, Rehovot 7610001, Israel\vspace*{2mm}}

\begin{abstract}
\vspace*{3mm}
A first measurement of time-reversal (T) asymmetries that are not also
CP asymmetries has been recently achieved by the \babar
collaboration. We analyze the measured asymmetries in the presence of
direct CP violation, CPT violation, wrong strangeness decays and wrong
sign semileptonic decays. We note that the commonly used $S_{\psi K}$ and $C_{\psi K}$ parameters are CP-odd, but have a T-odd CPT-even part and a T-even CPT-odd part. We introduce parameters that have well-defined transformation properties under CP, T and CPT. We identify contributions to the measured asymmetries that are T conserving.
We explain why, in order that the measured asymmetries would be purely odd under time-reversal, there is no need to assume the absence of direct CP violation. Instead, one needs to assume (i) the absence of CPT violation in strangeness changing decays, and (ii) the absence of wrong sign decays.
\end{abstract}

\maketitle

\section{Introduction}
\label{sec:intro}
The \babar collaboration has recently announced the first direct observation of time-reversal violation in the neutral $B$ meson system~\cite{Lees:2012uka}. The basic idea is to compare the time-dependent rates of two processes that differ by exchange of initial and final states. The measurement makes use of the Einstein-Podolsky-Rosen (EPR) effect in the entangled $B$ mesons produced in $\Upsilon(4S)$ decays~\cite{Banuls:1999aj,Wolfenstein:1999re,Bernabeu:2012ab,Alvarez:2006nk,Bernabeu:2013qea}. For example, one rate, $\Gamma_{(\psi K_L)_\perp,\ell^-X}$, involves the decay of one of the neutral $B$'s into a $\psi K_L$ state, and, after time $t$, the decay of the other $B$ into $\ell^- X$. The other rate, $\Gamma_{(\ell^+ X)_\perp,\psi K_S}$, involves the decay of one of the neutral $B$'s into $\ell^+ X$, and, after time $t$, the decay of the other $B$ into $\psi K_S$. Under certain assumptions, to be spelled out below, this is a comparison between the rates of $B_-\to\overline{B}{}^0$ and $\overline{B}{}^0\to B_-$, where $\overline{B}{}^0$ has a well-defined flavor content ($b\bar d$) and $B_-$ is a CP-odd state. The asymmetry is defined as follows:
\beq\label{eq:atbexp}
A_{T}\equiv
\frac{\Gamma_{(\psi K_L)_\perp,\ell^- X}-\Gamma_{(\ell^+ X)_\perp,\psi K_S}}
{\Gamma_{(\psi K_L)_\perp,\ell^- X}+\Gamma_{(\ell^+ X)_\perp,\psi K_S}}\,.
\eeq
\babar take the time dependence of the asymmetry to be of the form
\beq\label{eq:aoft}
A_{T}=\frac{\Delta S_T^+}{2}\sin(\Delta m t)+\frac{\Delta C_T^+}{2}\cos(\Delta m t)\,.
\eeq
(A more accurate description of the \babar analysis is given in Section~\ref{sec:tra}.)
They obtain
\beq\label{eq:stexp}
\Delta S_T^+=-1.37\pm0.15\,,\qquad\Delta C_T^+=+0.10\pm0.16\,.
\eeq
The fact that $\Delta S_T^+\neq0$ constitutes their direct demonstration of time reversal violation.

Time reversal violation had been observed earlier in the neutral $K$
meson system by the CPLEAR collaboration~\cite{Angelopoulos:1998dv}. The measurement involves the processes $p\bar p\to K^-\pi^+ K^0$ and $p\bar p\to K^+\pi^-\overline{K}{}^0$. Again, one aims to compare rates of processes that are related by exchange of initial and final states. One rate, $\Gamma_{K^-,e^-}$, involves a production of $K^-$ and a neutral $K$ that after time $t$ decay into $e^-\pi^+\bar\nu$. The other rate, $\Gamma_{K^+,\ell^+}$, involves the production of $K^+$ and a neutral $\overline{K}$ that after time $t$ decay  into $e^+\pi^-\nu$. Under certain assumptions, this is a comparison between the rates of $K^0\to\overline{K}{}^0$ and $\overline{K}{}^0\to K^0$~\cite{Kabir:1970ts}. The time dependent asymmetry is defined as follows:
\beq
A_{T,K}\equiv
\frac{\Gamma_{K^+,e^+}-\Gamma_{K^-,e^-}}
{\Gamma_{K^+,e^+}+\Gamma_{K^-,e^-}}\,.
\eeq
CPLEAR integrate the rates over times $\tau_S\leq t\leq20\tau_S$ ($\tau_S$ is the lifetime of $K_S$), and obtain
\beqa
\langle A_{T,K}\rangle_{(1-20)\tau_S}=(6.6\pm1.6)\times10^{-3}\,.
\eeqa
The CPLEAR asymmetry is also a CP asymmetry, since the initial and final states are CP-conjugate. In contrast, the \babar asymmetry is not a CP asymmetry.

In this work, we examine the analysis of $A_{T}$, with the aim of answering the following two related questions:
\begin{itemize}
\item Would it vanish in the T-symmetry limit?
\item Is the initial state in each of the two processes the T-conjugate of the final state in the other?
\end{itemize}
To answer these questions, it is helpful to use only parameters that have well-defined transformation properties under all three relevant symmetries: CP, T and CPT. In most of the related literature, CPT conservation is assumed, and then parameters that are CP-odd or CP-even are used. For example, the parameters $\Delta S_T^+$ and $\Delta C_T^+$ of Eq.~(\ref{eq:aoft}) are both CP-odd. However, when aiming to demonstrate that time-reversal is violated, one needs to allow for CPT violation~\cite{Schubert:1970,Lavoura:1990cg}. (Otherwise, T violation follows from CP violation.) In this case, most of the parameters used in the literature do not have well-defined transformation under T and CPT. In particular, $\Delta S_T^+$ and $\Delta C_T^+$ have, apart from a T-odd CPT-even part, also a T-even CPT-odd part.

In Section~\ref{sec:notations} we present our formalism and, in
particular, introduce parameters with well-defined transformation
properties under CP, T and CPT. In Section~\ref{sec:tra} we derive
expressions for the asymmetries measured by \babar in terms of these
parameters. The results allow us to answer the first question and to clearly
formulate the assumptions one needs to make in order to identify
the asymmetries measured by \babar with time reversal violation. In
Section~\ref{sec:time} we comment on the time-dependence of the
asymmetry under various approximations.
As concerns the second question, for two processes to be related by time-reversal, the initial state in each of them should be the time-reversal conjugate of the final state in the other. The subtleties related to this requirement, in the context of the \babar measurements, are clarified in Section~\ref{sec:direct}. We conclude in Section~\ref{sec:con}. In Appendix~\ref{sec:data} we compile relevant experimental constraints on CP violation in mixing and in decay, on CPT violation, and on wrong-strangeness and wrong-sign $B$ decays. In Appendix~\ref{sec:propose} we present combinations of measurements that allow us to isolate various parameters of interest. Appendix~\ref{sec:rates} contains the full expressions for all the asymmetries measured at \babar. In Appendix~\ref{sec:ratest} we define ``theoretical asymmetries" (or, equivalently, gedanken experiments) involving inverse decays, which provide another way to shed light on the subtleties of experimentally demonstrating time reversal violation. In Appendix~\ref{sec:EPR} we show how CPT violation affects the EPR entanglement of the initial $B$ meson state.

\section{Definitions and Notations}
\label{sec:notations}
We use a formalism that allows for CPT violation. Similar to Ref.~\cite{AlvarezGaume:1998yr}, we use {\it in} and {\it out} states defined by the reciprocal basis~\cite{Silva:2000db}. Translating our notations to those of Ref.~\cite{Branco:1999fs} is straightforward.

\subsection{$B^0-\overline{B}{}^0$ mixing}
The effective Hamiltonian $H=M-i\Gamma/2$ describing the evolution of the oscillating system is non-Hermitian. It is thus diagonalized by a general similarity transformation,
\beq
X^{-1}HX={\rm diag}(\omega_H,\omega_L)\,,
\eeq
where
\beq\label{eq:defpqz}
X=\begin{pmatrix} p\sqrt{1+z}\ &\ p\sqrt{1-z} \\ -q\sqrt{1-z}\ & \ q\sqrt{1+z}\end{pmatrix}\,,
\eeq
with $|p|^2+|q|^2=1$. The complex parameter $z\equiv z^R+iz^I$ represents CP and CPT violation in mixing. We define a real parameter, $R_M$, representing T and CP violation in mixing:
\beq
R_M\equiv \frac{1-\abs{q/p}^2}{1+\abs{q/p}^2}\,.
\eeq
In Eq.~(\ref{eq:defpqz}) we omitted normalization factors that deviate from unity only in ${\cal O}(R_Mz,z^2)$.

For the mass and width eigenvalues, we use
\beqa
&&\omega_{H,L}=m_{H,L}-\frac i2\Gamma_{H,L}\,,\qquad
m=\frac{m_L+m_H}{2} \, , \qquad
\Gamma=\frac{\Gamma_L+\Gamma_H}{2}\,,\no\\
&&x=\frac{m_H-m_L}{\Gamma}\,,\qquad
y=\frac{\Gamma_H-\Gamma_L}{2\Gamma}\,.
\eeqa
The incoming mass eigenstates are
\beqa
|B_H^{\rm in}\rangle&=&
p\sqrt{1+z}|B^0\rangle-q\sqrt{1-z}|\overline{B}{}^0\rangle\,, \nonumber\\
|B_L^{\rm in}\rangle&=&
p\sqrt{1-z}|B^0\rangle+q\sqrt{1+z}|\overline{B}{}^0\rangle\,.
\eeqa
Their time evolution is given by
\beq
|B_{H,L}^{\rm in}(t)\rangle=e^{-i\omega_{H,L}t}|B_{H,L}^{\rm in}\rangle\,.
\eeq
The outgoing states are
\beqa
\!\!\!\!\!\!\!\!
\langle B_H^{\rm out}|&=&\frac{1}{2pq}(q\sqrt{1+z}\langle
B^0|-p\sqrt{1-z}\langle\overline{B}{}^0|)\,, \nonumber\\
\langle B_L^{\rm out}|&=&\frac{1}{2pq}(q\sqrt{1-z}\langle B^0|+p\sqrt{1+z}\langle\overline{B}{}^0|)\,.
\eeqa
%

\subsection{$B$ decay amplitudes}
We define decay amplitudes,
\beq
A_f\equiv A(B^0\to f)=\langle f|T|B^0\rangle\,,\qquad
\bar A_f\equiv A(\overline{B}{}^0\to f)=\langle f|T|\overline{B}{}^0\rangle\,,
\eeq
and the inverse-decay amplitudes,
\beq
A_f^{\rm ID}\equiv A(f^T\to B^0)=\langle B^0|T|f^T\rangle\,,\qquad
\bar A_f^{\rm ID}\equiv A(f^T\to \overline{B}{}^0)=\langle\overline{B}{}^0|T|f^T\rangle\,,
\eeq
where $f^T$ is T-conjugate ({\it i.e.} reversed spins and momenta) to $f$.
CP violation in decay and in inverse decay is manifest in the direct asymmetries
\beq\label{eq:adf}
{\cal A}_f\equiv\frac{|\bar A_{\bar f}|^2-|A_{f}|^2}{|\bar A_{\bar f}|^2+|A_{f}|^2}\,, \qquad
{\cal A}^{\rm ID}_f\equiv\frac{|A^{\rm ID}_{f}|^2  - |\bar A^{\rm ID}_{\bar f}|^2}{|A^{\rm ID}_{f}|^2+|\bar A^{\rm ID}_{\bar f}|^2}\,,
\eeq
where $\bar f$ is CP-conjugate to $f$.

We define complex parameters, $\theta_{f}$, representing CPT violation in the decay:
\beq
	\theta_{f} = \theta^R_{f}+i\theta^I_{f}
\equiv 	\frac{A^{\rm ID}_{\bar f}/\bar A^{\rm ID}_{\bar f}  -  \bar A_{ f}/A_{ f}  }{A^{\rm ID}_{\bar f}/\bar A^{\rm ID}_{\bar f}  +  \bar A_{ f}/A_{ f}}  \, .
\eeq
Under CP, $\theta_{f} \to - \theta_{\bar f} $, while under T, $\theta_{f} \to \theta_{\bar f} $. Thus, for final CP eigenstates, $\theta_{f}\neq0$ breaks CPT and CP, but not T.

We further define the phase convention independent combination of amplitudes and mixing parameters,
\begin{align}\label{eq:deflamf}
	\lambda_f \equiv \frac{q}{p}\frac{\bar A_f}{A_f} \, \sqrt{\frac{1+\theta_{ f}}{1-\theta_{f}}}
= \frac{q}{p}\frac{A^{\rm ID}_{\bar f}}{\bar A^{\rm ID}_{\bar f}}\, \sqrt{\frac{1-\theta_{f}}{1+\theta_{f}}} \, .
\end{align}
In the CPT limit $z=\theta_{f}=0$ and the standard definition of $\lambda_f$ is recovered. It is convenient to introduce the following functions of $\lambda_f$:
\beq\label{eq:defcsg}
C_f\equiv\frac{1-|\lambda_f|^2}{1+|\lambda_f|^2}\,,\qquad
S_f\equiv\frac{2\,{\cal I}m(\lambda_f)}{1+|\lambda_f|^2}\,,\qquad
G_f\equiv\frac{2\,{\cal R}e(\lambda_f)}{1+|\lambda_f|^2}\,,
\eeq
with $C^2_f+G^2_f+S^2_f=1$.

We emphasize that the definition of $\lambda_f$ in Eq.~(\ref{eq:deflamf}) and, consequently, the definitions of $S_f$, $C_f$ and $G_f$ in Eq.~(\ref{eq:defcsg}) differ from the standard definitions of these parameters in the literature. Our definition lends itself straightforwardly to the theoretical analysis that we are doing. The standard definition lends itself straightforwardly to the description of the experimentally measured rates. The relation between the two will be further clarified in the next subsection. The distinction between the two is crucial for understanding the subtleties in the interpretation of the \babar measurements. Of course, in the absence of CPT violation, the two sets of definitions coincide.

\subsection{T, CP and CPT transformations}
The transformation rules for the parameters defined in the previous subsection under T, CP and CPT, are summarized in Table~\ref{tab:tra}.

\begin{table}[h!]
\begin{center}
\begin{tabular}{|c|c|c|c|c|c|c|c|c|} \hline\hline
\rule{0pt}{1.2em}%
~~Parameter~~   & T      & CP       & CPT  \cr \hline\hline
$R_M$       & $-R_M$ & $-R_{M}$ & $R_M$  \cr
$z$         & $z$    & $-z$     & $-z$   \cr
$\lambda_f$ & $1/\lambda_{\bar f}$ & $1/\lambda_{\bar f}$ & $\lambda_f$ \cr
$S_{f}$     & $-S_{\bar f}$ & $-S_{\bar f}$ & $S_{f}$ \cr
$C_{f}$     & $-C_{\bar f}$ & $-C_{\bar f}$ & $C_{f}$ \cr
$G_{f}$     & ~~$G_{\bar f}$~~  & ~~$G_{\bar f}$~~  & ~~$G_{f}$~~ \cr
${\cal A}_f$ & $-{\cal A}^{\rm ID}_f$ & $-{\cal A}_f$ & ${\cal A}^{\rm ID}_f$  \cr
$\theta_{f}$   & $\theta_{\bar f}$ & $-\theta_{\bar f}$ & $-\theta_{f}$ \\[3pt]
\hline\hline
\end{tabular}
\end{center}
\caption{Transformation rules of the various parameters under T, CP and CPT}
\label{tab:tra}
\end{table}

As explained above, it is very convenient for our purposes to use parameters that are either even or odd under all three transformations. Using superscript $+$ for T-even, and $-$ for T-odd, we define the following combinations:
\beqa\label{eq:DefLeptons}
C_f^-&=&\frac{1}{2}(C_f+C_{\bar f})\,,\;\;\;\;\;\;\;\;\;\;\;\;\;C_f^+=\frac{1}{2}(C_f-C_{\bar f})\,,\no\\
S_f^-&=&\frac{1}{2}(S_f+S_{\bar f})\,,\;\;\;\;\;\;\;\;\;\;\;\;\;\;S_f^+=\frac{1}{2}(S_f-S_{\bar f})\,,\no\\
G_f^-&=&\frac{1}{2}(G_f-G_{\bar f})\,,\;\;\;\;\;\;\;\;\;\;\;\;\;G_f^+=\frac{1}{2}(G_f+G_{\bar f})\,,\no\\
\theta_{f}^-&=&\frac{1}{2}(\theta_{f}-\theta_{\bar f})\,,\;\;\;\;\;\;\;\;\;\;\;\;\;\;\;\;\theta_{f}^+=\frac{1}{2}(\theta_{f}+\theta_{\bar f})\,.
\eeqa
We further define
\beq\label{eq:apm}
{\cal A}^-_{f}=\frac{{\cal A}_f+{\cal A}^{{\rm ID}}_{f}}{2}=R_M-C_{f_{CP}}\,,\qquad {\cal A}^+_{f}=\frac{{\cal A}_{f}-{\cal A}^{{\rm ID}}_{f}}{2}=-\theta_{f_{CP}}^{+,R}\,,
\eeq
where  the last step in each equation is relevant only for CP eigenstates and we expand to linear order in $C_{f_{CP}}$, $R_M$ and $\theta_{f_{CP}}$.
A summary of the transformation properties of these parameters is
provided in Table~\ref{tab:exp}.
For final CP eigenstates, such as $f=\psi K_{S,L}$, $S_f$ and $C_f$
are T-odd, while $G_f$ and $\theta_f$ are T-even, so that
$S_f^+=C_f^+=G_f^-=\theta_f^-=0$. We summarize the transformation
properties for final CP eigenstates also in Table \ref{tab:exp}.

\begin{table}[t!]
\begin{center}
\begin{tabular}{|c|c|c|c|} \hline\hline
\rule{0pt}{1.2em}%
Parameter        & T  &CP& CPT  \cr \hline\hline
~~$R_M,~S_f^-,~C_f^-,~G_f^-,~{\cal A}^-_f$~~          & ~~~$-$~~~  & ~~~$-$~~~  &  ~~~$+$~~~  \cr
$z,~\theta_f^+~,{\cal A}_f^+$           & $+$  & $-$  &  $-$  \\[3pt]
$\theta_{f}^-$   & $-$  & $+$  &  $-$   \\[3pt]
$S_f^+~,C_f^+~,G_f^+$          & $+$  & $+$  &  $+$   \\[3pt]
\hline
$S_{f_{CP}},C_{f_{CP}}$     & $-$  & $-$  &  $+$    \\[3pt]
$\theta_{f_{CP}}$           & $+$  & $-$  &  $-$    \\[3pt]
$G_{f_{CP}}$                & $+$  & $+$  &  $+$    \\[3pt]
\hline\hline
\end{tabular}
\end{center}
\caption{Eigenvalues of the various parameters under T, CP and CPT}
\label{tab:exp}
\end{table}

In practice, inverse decays are not accessible to the experiments. In particular, experiments are not sensitive to $\lambda_f$, as defined in Eq.~(\ref{eq:deflamf}), but to the related observable $\lambda_f^e$, defined via
\beq
\lambda_f^e\equiv \frac{q}{p}\frac{\bar A_f}{A_f}=\lambda_f\left(1-\theta_{f}\right)\,,
\eeq
where the second equation holds to first order in $\theta_f$. Accordingly, the experiments are sensitive to the following parameters:
\beqa\label{eq:cgse}
C_{f}^e&=&C_f+(1-C^2_f)\theta_{f}^R\,, \no\\
S_{f}^e&=&S_f(1-C_f\theta^R_f)-G_f\theta_{f}^I\,, \no\\
G_{f}^e&=&G_f(1-C_f\theta^R_f)+S_f\theta_{f}^I\,.
\eeqa
%

\subsection{Wrong-strangeness and wrong-sign decays}
Among the final CP eigenstates, we focus on decays into the final $\psi K_{S,L}$ states (neglecting effects of $\epsilon_K$). We distinguish between the right strangeness decays,
\beq
A_{K^0}=A(B^0\to K^0)\,,\qquad
\bar A_{\overline{K}{}^0}=A(\overline{B}{}^0\to\overline{K}{}^0)\,,
\eeq
and the wrong strangeness decays,
\beq
\bar A_{K^0}=A(\overline{B}{}^0\to K^0)\,,\qquad
A_{\overline{K}{}^0}=A(B^0\to\overline{K}{}^0)\,.
\eeq
We define
\beqa
\hat C_{\psi K}\equiv\frac12(C_{\psi K_S}+C_{\psi K_L})\,,&\qquad
&\Delta C_{\psi K}\equiv\frac12(C_{\psi K_S}-C_{\psi K_L})\,,\no\\
\hat S_{\psi K}\equiv\frac12(S_{\psi K_S}-S_{\psi K_L})\,,&\qquad
&\Delta S_{\psi K}\equiv\frac12(S_{\psi K_S}+S_{\psi K_L})\,,\no\\
\hat G_{\psi K}\equiv\frac12(G_{\psi K_S}-G_{\psi K_L})\,,&\qquad
&\Delta G_{\psi K}\equiv\frac12(G_{\psi K_S}+G_{\psi K_L})\,,
\eeqa
and
\beq
\hat\theta_{\psi K}\equiv\frac12(\theta_{\psi K_L}+\theta_{\psi K_S})\,,\qquad
\Delta\theta_{\psi K}\equiv\frac12(\theta_{\psi K_S}-\theta_{\psi K_L})\,.
\eeq
In the limit of no wrong strangeness decay, $\lambda_{\psi K_S}=-\lambda_{\psi K_L}$~\cite{Grossman:2002bu} (Ref.~\cite{Grossman:2002bu} assumes CPT conservation, but this is not a necessary assumption) and, consequently, $\Delta C_{\psi K},\Delta G_{\psi K},\Delta S_{\psi K}$ and $\Delta \theta_{\psi K}$ vanish.

Among the flavor specific final states, we focus on decays into final $\ell^\pm X$ states. Here we distinguish between the right sign decays,
\beq
A_{\ell^+}=A(B^0\to\ell^+ X)\,,\qquad
\bar A_{\ell^-}=A(\overline{B}{}^0\to\ell^- X)\,,
\eeq
and the wrong sign decays,
\beq
A_{\ell^-}=A(B^0\to\ell^- X)\,,\qquad
\bar A_{\ell^+}=A(\overline{B}{}^0\to\ell^+ X)\,.
\eeq
We define $C_{\ell}^\pm,S_{\ell}^\pm$ and $G_{\ell}^\pm$ according to Eq.~\eqref{eq:DefLeptons}, with $f=\ell^+$, and a super-index $+$ ($-$) denoting a T conserving (violating) combination.
Taking the wrong sign decays to be much smaller in magnitude than the right sign decays, we have $|\lambda_{\ell^+}|\ll1$ and $|\lambda_{\ell^-}^{-1}|\ll1$. We will work to first order in $|\lambda_{\ell^+}|$ and in $|\lambda_{\ell^-}^{-1}|$, which means that we set
$C_\ell^+=1$ and $C_\ell^-=0$.
On the other hand, the four other relevant parameters are linear in $|\lambda_{\ell^+}|$ and in $|\lambda_{\ell^-}^{-1}|$:
\beq
S_{\ell}^\pm\simeq{\cal I}m(\lambda_{\ell^+}\pm\lambda_{\ell^-}^{-1})\,,\qquad
G_{\ell}^\pm\simeq{\cal R}e(\lambda_{\ell^+}\pm\lambda_{\ell^-}^{-1})\,.
\eeq
%

\section{Time-reversal asymmetries}
\label{sec:tra}
\subsection{The master formula}
Consider a pair of $B$-mesons produced in $\Upsilon(4S)$ decay, where one of the $B$-mesons decays at time $t_1$ to a final state $f_1$, and the other $B$-meson decays at a later time, $t_2=t_1+t$, to a final state $f_2$. The time dependent rate for this process, to linear order in $R_M,z$ and $\theta$, is given by
\beqa\label{eq:master}
\Ghg{f_1}{f_2}&=&N_{(1)_\perp,2}e^{-\Gamma(t_1+t_2)}\\
&\times&\left[
\kappa_{(1)_\perp,2}\cosh(y\Gamma t)
+\sigma_{(1)_\perp,2}\sinh(y\Gamma t)
+C_{(1)_\perp,2}\cos(x\Gamma t)
+S_{(1)_\perp,2}\sin(x\Gamma t)\right].\nonumber
\eeqa
where
\beqa
N_{(1)_\perp,2}&=&\left[ 1+ (C_1 + C_2)(R_M-z^R) \right]{\cal N}_{1}{\cal N}_{2}\,,\nonumber \\
\kappa_{(1)_\perp,2}&=&	\Big[
	1-G_1 G_2
	+ (C_1+C_2+C_2G_1+C_1G_2-C_2 G_2 G_1 - C_1 G_2 G_1)z^R \nonumber\\
	&&~~~~- (S_1+S_2)z^I
	+G_1 G_2 (C_2\theta_2^R+C_1 \theta_1^R)
	-G_1 S_2\theta_2^I	-G_2 S_1\theta_1^I \Big]\,, \nonumber\\
\sigma_{(1)_\perp,2}&=&\Big[
	G_2-G_1-(C_2 - C_1 - C_2 G_2 - C_1 G_2 + C_2 G_1 + C_1 G_1)z^R\nonumber\\
	&&~~~~-(G_2 S_1-G_1 S_2)z^I
	-C_2 G_2\theta^R_2+ S_2\theta^I_2+C_1 G_1\theta^R_1- S_1\theta^I_1
	\Big]\,,\nonumber\\
C_{(1)_\perp,2}&=&-\Big[ C_2 C_1 + S_2 S_1
  +(C_2^2 C_1 + C_2 C_1^2 + C_1 G_2 + C_2 G_1 + C_2 S_2 S_1 + C_1 S_2
  S_1) z^R  \nonumber\\
&&~~~~~- (S_2 + S_1) z^I -G_2 S_1 \theta_2^I + (C_1 - C_2^2 C_1 - C_2
S_2 S_1) \theta_2^R
\nonumber\\ &&~~~~~~~~~~-
 G_1 S_2 \theta_1^I
 + (C_2 - C_2 C_1^2 - C_1 S_2 S_1) \theta_1^R
  \Big]\,, \nonumber\\
S_{(1)_\perp,2}&=&\Big[
	C_1 S_2 - C_2 S_1
  +(C_2 C_1 S_2 + C_1^2 S_2 + G_1 S_2 - C_2^2 S_1 - C_2 C_1 S_1 - G_2 S_1) z^R
 \nonumber\\
&& ~~~~~
+ (C_2-C_1 ) z^I  -  C_1 G_2 \theta_2^I
+ (C_2^2 S_1-C_2 C_1 S_2 - S_1 ) \theta_2^R
\nonumber\\ &&~~~~~~~~~~
+  C_2 G_1 \theta_1^I
 - (C_1^2 S_2 - C_2 C_1 S_1-S_2) \theta_1^R
 \Big]\,,
\eeqa
and, for the sake of brevity, we denote $X_i\equiv X_{f_i}$ for $X=S,C,G,\theta$ and ${\cal N}_{i}=\left( |A_{f_i}|^2 + |\bar{A}_{f_i}|^2 \right)$.
Eq.~(\ref{eq:master}) is our ``master formula", and serves as the starting point for all our calculations. Note that the decomposition into $e^{-\Gamma(t_1+t_2)}\times f(t)$ for the $\Upsilon(4S)$ decay products holds even in the presence of CPT violation. (See Appendix~\ref{sec:EPR}.)

\subsection{The \babar T-asymmetry}\label{sec:babar}
In what follows we set $y=0$, $C_\ell^+=1$, $C_\ell^-=0$ and $\epsilon_K=0$. We do so for the sake of simplicity: All these effects can, in principle, be accounted for. (Care is needed when accounting for the effect of Kaon mixing~\cite{Azimov:1980,Azimov:1998sz,Lipkin:1999qz,Bigi:2005ts,Calderon:2007rg,Grossman:2011zk}.) We work to linear order in the following small parameters:
\beq\label{eq:small}
R_M,\ z^R,\ z^I,\ \theta^R_f,\ \theta^I_f,\ \hat C_{\psi K}, \ \Delta C_{\psi K},\ \Delta S_{\psi K},\ \Delta G_{\psi K}, S_\ell^\pm,\ G_\ell^\pm.
\eeq

We consider the two processes that are relevant to the experimentally measured asymmetry~(\ref{eq:atbexp}). Using the master formulas~(\ref{eq:master}), and implementing our approximations, we obtain the following time-dependent decay rates:
\beqa\label{eq:lmrate}
	\Ghg{\psi K_L}{\ell^-X}&=& \left[ 1- R_M+z^R \right] {\cal N}_{\psi K_L}{\cal N}_{\ell^-X}\nonumber\\
&\times&	\Bigg\{
	\Big[
	1-G_{\psi K_L} G_{\ell^-}- (1+G_{\psi K_L})z^R
	- S_{\psi K_L}z^I \Big]\nonumber\\
&	-&\Big[ -C_{\psi K_L} + S_{\ell^-} S_{\psi K_L}
  -G_{\psi K_L} z^R  -  S_{\psi K_L} z^I - \theta_{\psi K_L}^R
  \Big] \cos(x\Gamma t)\nonumber\\
& +&\Big[ S_{\psi K_L} - S_{\psi K_L} z^R - z^I
- G_{\psi K_L} \theta_{\psi K_L}^I
 \Big] \sin(x\Gamma t)
	\Bigg\} \, ,\\
	\Ghg{\ell^+X}{\psi K_S}&=& \left[ 1+ R_M-z^R \right]{\cal
          N}_{\ell^+X}{\cal N}_{\psi K_S} \label{eq:psrate}\nonumber \\
&\times& \Bigg\{
	\Big[
	1-G_{\ell^+} G_{\psi K_S}
	+ (1+G_{\psi K_S})z^R
	- S_{\psi K_S}z^I\Big]\nonumber\\
&	-&\Big[ C_{\psi K_S} + S_{\psi K_S} S_{\ell^+}
  +G_{\psi K_S} z^R  - S_{\psi K_S} z^I +  \theta_{\psi K_S}^R
  \Big] \cos(x\Gamma t)\nonumber\\
&	+&\Big[
	S_{\psi K_S} +S_{\psi K_S} z^R - z^I
- G_{\psi K_S} \theta_{\psi K_S}^I
 \Big] \sin(x\Gamma t)
	\Bigg\} \, ,
\eeqa
where the overall decay exponential factor is omitted.
The analysis performed by \babar, as described in Ref.~\cite{Lees:2012uka}, is as follows. The time dependent decay rates are measured and fitted to time-dependence of the form~(\ref{eq:master}), approximating (as we do) $y=0$. The quantities $\Delta S_T^+$ and $\Delta C_T^+$, to which the \babar results of (\ref{eq:stexp}) apply, correspond to the following combinations:
\beqa\label{eq:defdstdct}
\Delta S_T^+&=&\frac{S_{(\psi K_L)_\perp,\ell^-X}}{\kappa_{(\psi K_L)_\perp,\ell^-X}}
-\frac{S_{(\ell^+ X)_\perp,\psi K_S}}{\kappa_{(\ell^+ X)_\perp,\psi K_S}}
\equiv S^-_{\ell^-,K_L}-S^+_{\ell^+,K_S}\,,\nonumber\\
\Delta C_T^+&=&\frac{C_{(\psi K_L)_\perp,\ell^-X}}{\kappa_{(\psi K_L)_\perp,\ell^-X}}
-\frac{C_{(\ell^+ X)_\perp,\psi K_S}}{\kappa_{(\ell^+ X)_\perp,\psi K_S}}
\equiv C^-_{\ell^-,K_L}-C^+_{\ell^+,K_S}\,.
\eeqa
The last identity relates our definitions in Eq.~(\ref{eq:master}) with those of \babar in Ref.~\cite{Lees:2012uka}. In the latter, a super-index $+$ refers to the case the leptonic tag occurs before the
CP tag, as in $\Ghg{\ell^+X}{\psi K_S}$, while a super-index $-$
refers to the case that the CP tag occurs before the leptonic tag, as
in $\Ghg{\psi K_L}{\ell^-X}$.
We note that the normalization of Eqs.~(\ref{eq:defdstdct}) removes the dependence on the total production rates and effects such as direct CP violation in leptonic decays.

We obtain the following expressions for these observables:
\beqa\label{eq:dstdct}
\Delta S_T^+&=&-2\left[(\hat S_{\psi K}-\hat G_{\psi K}\hat\theta_{\psi K}^I)
+\hat S_{\psi K}\hat G_{\psi K}(G_\ell^- - z^R)\right]\,,\nonumber\\
\Delta C_T^+&=&2\left[(\hat C_{\psi K}+\hat\theta_{\psi K}^R)+\hat S_{\psi K}(S_\ell^- - z^I)\right]\,.
\eeqa
If we switch off all the T-odd parameters, we are left with the following T conserving (TC) contributions:
\beqa\label{eq:tozero}
\Delta S_T^+(\mbox{T-odd parameters}=0)&=&2\hat G_{\psi K}\hat\theta_{\psi K}^I\,,\nonumber\\
\Delta C_T^+(\mbox{T-odd parameters}=0)&=&2\hat\theta_{\psi K}^R\,.
\eeqa
These contributions are CPT violating. Yet, since $\hat\theta_{\psi K}$ involves inverse decays, we are not aware of any way to experimentally bound it, and to exclude the possibility that it generates the measured value of $\Delta S_T^+$.
We would like to emphasize, however, the following three points.
\begin{itemize}
\item The appearance of the T conserving, CPT violating effects should come as no surprise. As explained in the discussion of Eq.~(\ref{eq:cgse}), experiments can only probe $S_{\psi K}^e$ and $C_{\psi K}^e$, which include these terms.
\item While we are not aware of any way to constrain $\hat\theta_{\psi K}$ from tree-level processes, it may contribute to measurable effects, such as CPT violation in mixing, at the loop level. In the absence of a rigorous framework that incorporates CPT violation, it is impossible to calculate such effects.
\item It would of course be extremely exciting if the \babar measurement is affected by CPT violating effects.
\end{itemize}

An additional interesting feature of Eqs.~(\ref{eq:dstdct}) is the appearance of terms that are quadratic in T-odd parameters,
\beqa\label{eq:toquad}
\Delta S_T^+(\mbox{quadratic in T-odd\ parameters})&=&-2\hat G_{\psi K}\hat S_{\psi K}G_\ell^-\,,\nonumber\\
\Delta C_T^+(\mbox{quadratic in T-odd parameters})&=&2\hat S_{\psi K}S_\ell^-\,.
\eeqa
While these terms would vanish if we take all T-odd parameters to
zero, they are still T-conserving. Note that since we expand to linear
order in all T-odd parameters, except for $\hat S_{\psi K_S}$, there
are additional T conserving, $\hat S_{\psi K}$-independent,
contributions that are quadratic in T-odd parameters that are not
presented in Eqs.~(\ref{eq:toquad}). Since $\hat G_{\psi K}^2+\hat
S_{\psi K}^2\le 1$, the maximal  absolute value of the term on the
right-hand side of Eq.~(\ref{eq:toquad}) for $\Delta S_T^+$ is 1,
$|2\hat G_{\psi K}\hat S_{\psi K}G_\ell^-|\le 1$. Thus, if experiments establish  $|\Delta S_T^+|>1$, such a result cannot be explained by this term alone.

We are now also able to formulate the conditions under which the \babar measurement would demonstrate T violation unambiguously:
\beq
\hat\theta_{\psi K}=G_\ell^-=S_\ell^-=0\,.
\eeqa
In words, the necessary conditions are the following:
\begin{itemize}
\item The absence of CPT violation in strangeness changing decays.
\item The absence of wrong sign decays or, if wrong sign decays occur, the absence of direct CP violation in semileptonic decays.
\end{itemize}

\subsection{The time dependence of $A_{T}$}
\label{sec:time}
It is often assumed that the time-dependence of the time-reversal asymmetry $A_{T}$ is of the form of Eq.~(\ref{eq:aoft}). Eqs.~(\ref{eq:lmrate}) and~(\ref{eq:psrate}) reveal, however, that even when taking $y=0$ and expanding to linear order in the various small parameters, the time dependence is more complicated:
\beq
A_{T}=R_T+C_T\cos(x\Gamma t)+S_T\sin(x\Gamma t)+B_T\sin^2(x\Gamma t)+D_T\sin(x\Gamma t)\cos(x\Gamma t)\,.
\eeq
Even when we neglect CPT violation, wrong sign decays, and wrong strangeness decays, the time dependence is, in general, more complicated than Eq.~(\ref{eq:aoft}). If, in addition, one neglects CP violation in decay, then the only source of T violating effects is in mixing and should therefore vanish at $t=0$:
\beq
A_{T}=-R_M[1-\cos(x\Gamma t)]-\hat S_{\psi K}\sin(x\Gamma t)+R_M\hat S_{\psi K}^2\sin^2(x\Gamma t)\,.
\eeq
One may argue that $R_M$ is experimentally constrained to be very small. But then one should also take into account the fact that $\hat C_{\psi K}$ is experimentally constrained to be very small, and the asymmetry has the simpler well-known form
\beq
A_{T}=-\hat S_{\psi K}\sin(x\Gamma t)\,.
\eeq
Moreover, the $\hat S_{\psi K}$ parameter is measured and known. The whole point of the \babar measurement is not to measure the values of the parameters, but to demonstrate time-reversal violation in processes that are not CP-conjugate. It is perhaps more appropriate not to take experimental information from previous measurements.

\section{CP violation in right-strangeness decays}
\label{sec:direct}
A-priori, one would expect that direct CP violation in right-strangeness decays is enough to allow for $A_T\neq0$ even in the T-symmetry limit. However, in the previous section we found that this is not the case. In this section we first explain the reasoning behind the naive expectation, and, second, obtain the conditions under which the two processes measured by \babar are not T conjugate.

In $\Gamma_{(\psi K_L)_\perp,\ell^-X}$, the initial $B$-meson state is orthogonal to the one that decays to $\psi K_L$. In $\Gamma_{(\ell^+X)_\perp,\psi K_S}$, the final state is the one that decays into $\psi K_S$. Are these two states identical? They would be if the state that does not decay to $\psi K_L$, $|B_{(\to\psi K_L)_\perp}\rangle$, and the state that does not decay into $\psi K_S$, $|B_{(\to\psi K_S)_\perp}\rangle$, were orthogonal to each other. Using
\beqa
|B_{(\to\psi K_L)_\perp}^{\rm in}\rangle&=&N_L\left[\bar A_{\psi K_L}|B^0\rangle-A_{\psi K_L}|\overline{B}{}^0\rangle\right]\,,\nonumber\\
|B_{(\to\psi K_S)_\perp}^{\rm in}\rangle&=&N_S\left[\bar A_{\psi
    K_S}|B^0\rangle-A_{\psi K_S}|\overline{B}{}^0\rangle\right]\,,
\eeqa
where $N_{S,L}$ are normalization coefficients, and neglecting wrong strangeness decays, we obtain
\beq \label{eq:prepoverlap}
\langle B_{(\to\psi K_S)_\perp}|B_{(\to\psi K_L)_\perp}\rangle\propto{\cal A}_{\psi K}\,,
\eeq
where ${\cal A}_f$ is the direct asymmetry defined in Eq.~(\ref{eq:adf}). (The same orthogonality condition can be obtained by using $\langle B_{\to\psi K_S}|B_{\to\psi K_L)}\rangle$.)
The conclusion is that the state that is orthogonal to the one that decays to $\psi K_L$ and the state that decays to $\psi K_S$ are not the same state, unless there is no direct CP violation in the $B\to\psi K$ decays. This is presumably the reason why the theoretical paper~\cite{Bernabeu:2012ab} and the experimental paper~\cite{Lees:2012uka} explicitly state that they neglect direct CP violation.

However, as we learned from the analysis in Section~\ref{sec:tra}, the correct question is not whether the state that does not decay to $\psi K_L$ is the same as the state that decays to $\psi K_S$. Instead, the question is whether it is the same as the state generated in the inverse decay of $\psi K_S$. (The orthogonality of the two $B$-mesons at $t_1$ is guaranteed by the EPR entanglement, allowing one to label the initial $B$ state at $t_1$.) This would be the case if $|B_{(\to\psi K_L)_\perp}\rangle$ and $|B_{(\psi K_S\to)_\perp}\rangle$ were orthogonal. Using
\beqa
|B_{\psi K_S\to}^{\rm in}\rangle&=&N_{S}^{\rm ID}\left[A^{\rm ID}_{\psi K_S}|B^0\rangle+\bar A^{\rm ID}_{\psi K_S}|\overline{B}{}^0\rangle\right]\,,\nonumber\\
|B_{(\psi K_S\to)_\perp}^{\rm in}\rangle&=&N_{S}^{\rm ID}\left[\bar A^{\rm ID*}_{\psi K_S}|B^0\rangle-A^{\rm ID*}_{\psi K_S}|\overline{B}{}^0\rangle\right]\,,
\eeqa
we obtain
\beq\label{eq:ol1}
\langle B_{(\psi K_S\to)_\perp}|B_{(\to\psi K_L)_\perp}\rangle
&&\propto \lambda_{\psi K_S}+\lambda_{\psi K_L}+( \lambda_{\psi K_S}-\lambda_{\psi K_L})\hat\theta_{\psi K}  \no\\
&&\propto \Delta C_{\psi K} - \frac{\Delta G_{\psi K}+i \Delta S_{\psi K}}{\hat{G}_{\psi K}+i\hat{S}_{\psi K}} -\hat\theta_{\psi K} \,.
\eeq
Similarly,
\beq\label{eq:ol1B}
\langle B_{(\psi K_L\to)_\perp}|B_{(\to\psi K_S)_\perp}\rangle
&&\propto \lambda_{\psi K_L}+\lambda_{\psi K_S}+( \lambda_{\psi K_L}-\lambda_{\psi K_S})\hat\theta_{\psi K}  \no\\
&&\propto \Delta C_{\psi K} - \frac{\Delta G_{\psi K}+i \Delta S_{\psi K}}{\hat{G}_{\psi K}+i\hat{S}_{\psi K}} +\hat\theta_{\psi K} \,.
\eeq
We thus learn that the state that is orthogonal to the one that decays to $\psi K_L(\psi K_S)$ is the same as the state that is generated in the inverse decay of $\psi K_S(\psi K_L)$, unless there are wrong strangeness decays or CPT violation in strangeness changing decays. In the absence of these phenomena, the two processes are related by exchange of the initial and final CP-tagged states, as required for time-reversal conjugate processes.

One can repeat an analogous analysis for the lepton-tagged states. The question to be asked is whether the state that does not decay to $\ell^+X$ is orthogonal to the state that is not generated in the inverse decay of $\ell^-X$. For the overlap between these two states, we obtain:
\beq\label{eq:ol2}
\langle B_{(\ell^- X\to)_\perp}|B_{(\to\ell^+ X)_\perp}\rangle\propto \lambda_{\ell^+}\,.
\eeq
If $\lambda_{\ell^+}=0$, the two states are orthogonal. We thus learn that the state that is orthogonal to the one that decays into $\ell^+ X$ is the same as the state that is generated in the inverse decay of $\ell^-X$, unless there are wrong sign decays and wrong sign inverse decays. If wrong sign decays can be neglected, then the two processes are related by exchange of the initial and final lepton-tagged states, as required for time-reversal conjugate processes. (For a related asymmetry, involving $\Gamma_{(\psi K_L)_\perp,\ell^+ X}$ and $\Gamma_{(\ell^- X)_\perp,\psi K_S}$, the corresponding condition is $1/\lambda_{\ell^-}=0$.)

If Eq.~\eqref{eq:ol1} and Eq.~\eqref{eq:ol2} are both zero, then $\Gamma_{(\psi K_L)_\perp, \ell^- X}=\Gamma_{(\psi K_S)^T,\ell^-X}$ and $\Gamma_{(\ell^+ X)_\perp, \psi K_S}=\Gamma_{(\ell^- X)^T,\psi K_S}$ (under proper normalization such that the number of initial $B$'s are equal). In this case, $A_T$ as defined in Eq.~(\ref{eq:atbexp}) is indeed a T-asymmetry.
We conclude that if wrong strangeness decays, wrong sign decays and CPT violation in strangeness changing decays can be neglected, then the two processes measured by \babar represent two T-conjugate processes, and then there should be no T conserving contributions to $\Delta S_T^+$ and $\Delta S_T^-$. In particular, one need not assume the absence of direct CP violation.

\section{Conclusions}
\label{sec:con}

The \babar collaboration has measured time-reversal asymmetries in $B$ decays.
Two main ingredients --- the EPR effect between the two $B$-mesons produced in
$\Upsilon(4S)$ decays and the availability of both lepton-tagging and CP-tagging ---
allow the experimenters to approximately realize the main principle of time-reversal conjugate
processes: exchanging initial and final states.

A precise exchange is impossible. The final state is identified by the $B$-meson decay,
and the T-conjugate process requires, therefore, that a $B$-meson is produced in the
corresponding inverse decay. Instead, the experimenters measure a process where the
initial $B$-meson is identified by the decay of the other, entangled $B$-meson. We showed, however, that the initial $B$-meson prepared by lepton tagging, and the one that would be
produced in the appropriate inverse decay are not identical only if there are wrong-sign
decays. The initial $B$-meson prepared by CP tagging, and the one that would be
produced in the appropriate inverse decay, are not identical only if there are
wrong-strangeness contributions, or in the presence of CPT violation in decays.

The effect of CPT violation in decay has gained very little attention in the literature. One reason is that it can only be probed by measuring both decay rates and inverse decay rates, but the latter are practically inaccessible to experiments. For precisely this reason, there are no bounds on these effects. In principle, they could play a significant role in the asymmetries measured by \babar, in spite of the fact that they are T conserving.

As concerns the questions posed in the Introduction, we find the following answers:
\begin{itemize}
\item The $A_T$ measurement by \babar would vanish in the T-symmetry limit provided that CPT is conserved in strangeness changing decays (see Eq.~\eqref{eq:tozero}).
\item The initial state in each of the two processes would be the T-conjugate of the final state in the other, provided that there are neither wrong strangeness decays nor wrong sign decays nor CPT violation in strangeness changing decays. In particular, there is no need to assume the absence of direct CP violation (see Eqs.~\eqref{eq:ol1}-\eqref{eq:ol2}).
\end{itemize}

Both wrong-sign and wrong-strangeness effects are expected to be very small. Yet, the existing experimental upper bounds on these
effects are rather weak. The same set of measurements used for the time-reversal
asymmetries can be used (in different combinations) to constrain also the
wrong-sign and wrong-strangeness contributions.

While in this work we concentrated on very specific measurements in
$B$ decays, our results are more general. They apply straightforwardly, with
minor changes, to other meson systems. The main ideas also apply to
neutrino oscillation experiments.
Observation of $P(\nu_\alpha \to \nu_\beta)\ne P(\nu_\beta \to \nu_\alpha)$ has been advocated as a way to establish T-violation.
Such a result can arise, however, also from nonstandard interactions in the production or the detection
processes~\cite{Grossman:1995wx,GonzalezGarcia:2001mp,Bellazzini:2010gn}.

\vspace{1cm}
\begin{center}

{\bf Acknowledgements}
\end{center}
We thank Jos\'e Bernab\'eu, Klaus Schubert and Abner Soffer for useful correspondence.
EA thanks the Kupcinet-Getz Program for their hospitality at the Weizmann Institute.
YG is a Weston Visiting Professor at the Weizmann Institute. This work was partially supported by a grant from the Simons Foundation ($\#$267432 to YG).
The work of YG is supported in part by the U.S. National Science Foundation through grant PHY-0757868 and by the United States-Israel Binational Science Foundation (BSF) under grant No.~2010221.
YN is the Amos de-Shalit chair of theoretical physics.
YN is supported by the Israel Science Foundation, by the I-CORE program of the
Planning and Budgeting Committee and the Israel Science Foundation (grant
number 1937/12), and by the
German-Israeli foundation for scientific research and development (GIF).

\appendix
\section{Relevant data}
\label{sec:data}
How certain is it that the small parameters listed in Eq.~(\ref{eq:small}) are indeed small? In this Appendix, we compile the relevant experimental constraints. We caution the reader that some of these constraints were derived making assumptions that we do not make. For example, CPT symmetry is assumed when deriving the bounds on $R_M$. Thus, the upper bounds and ranges that we quote below assume no cancelations among different contributions and, even then, should be taken only as rough estimates.

Based on Ref.~\cite{Amhis:2012bh}, we find the following range for $R_M$:
\begin{align}
&	R_M = (-0.2\pm2.0) \times 10^{-3}\quad \text{from } \Upsilon({\rm 4S}) \, , \\
&	R_M = (+0.3\pm0.9) \times 10^{-3} \quad \text{world average }  \, . 	
\end{align}
Ref.~\cite{Beringer:1900zz} uses the \babar measurement
\cite{Aubert:2004xga} to constrain $z$:
\begin{align}
&	z^R = \left( -2\pm5 \right)\times 10^{-2} \,  , \quad
	z^I = \left( +0.8\pm0.4 \right)\times 10^{-2}	 \, .
\end{align}
Ref.~\cite{Amhis:2012bh} obtains the following ranges for the $S_{\psi K_{S,L}}$ and $C_{\psi K_{S,L}}$ parameters:
\begin{align} \label{eq:HFAGS}
&	S_{\psi K_S}=+0.665\pm0.024 \, , \qquad
	C_{\psi K_S} =+0.024\pm0.026 \, , \nonumber\\
&	S_{\psi K_L}= - 0.663\pm0.041 \, , \qquad	
	C_{\psi K_L} =-0.023\pm0.030 \, .
\end{align}
Note that here we take $S_{\psi K_L}\to - S_{\psi K_L}$ with respect to \cite{Amhis:2012bh}. Naively combining these ranges, we obtain
\begin{align}
&	\Delta C_{\psi K} = +0.02\pm 0.02 \, , \qquad
	\hat{C}_{\psi K} = +0.00\pm 0.02 \, ,  \\
&	\Delta S_{\psi K} = +0.00\pm 0.02 \, ,  \qquad
	\hat{S}_{\psi K}	= +0.66\pm 0.02 \, . \label{eq:SpsiK}
\end{align}
Direct bounds on wrong-strangeness $B$ decays are given by the \babar collaboration in Ref.~\cite{Aubert:2004ei}. Ref.~\cite{Beringer:1900zz} quotes $|\lambda_{\psi K^{*0}}| < 0.25\ {\rm at}\ 95\%\ {\rm CL}$, which is weaker than the above results.

As concerns wrong-sign $B$ decays, we use the individual branching ratios from Ref.~\cite{Beringer:1900zz} to obtain
\begin{align}\label{eq:wsrsexp}
 \frac{{\rm BR}(B^+\to \ell^+ \nu X)  }{ {\rm BR} (B^0 \to \ell^+\nu X) }
 =  1.06\pm0.04 \, .
\end{align}
In the isospin limit and in the absence of wrong-sign decays, we should have
\begin{align}\label{eq:wsrsthe}
\frac{{\rm BR}(B^+\to \ell^+ \nu X)  }{ {\rm BR} (B^0 \to \ell^+\nu X) }
 = \frac{\tau_{B^+}}{\tau_{B^0}} = 1.08\pm0.01 \, .
\end{align}
Comparing Eq. (\ref{eq:wsrsexp}) to (\ref{eq:wsrsthe}) we obtain, at $2\sigma$
\begin{align}
\left|{\bar A_{\ell^+}/A_{\ell^+}}\right| < 0.44 \, .
\end{align}
Using the exclusive process $B^+\to \ell^+ \nu \overline{D}{}^0$ instead of the inclusive one results in a weaker bound. One can also use different tagging techniques in measurements of CP asymmetries to place bounds on wrong sign decays. Assuming CPT to be a good symmetry, we find
\begin{align}
	C^{\rm lep}_{CP} - C^{\rm Kaon}_{CP} = \hat{S}_{\psi K_S} {S}_\ell^+ \, ,
\end{align}
where $C^{\rm lep}_{CP}$ and $C^{\rm Kaon}_{CP}$ are the measured coefficients for the $\cos(x\Gamma t)$ function with lepton and Kaon tagging, respectively. Using the results of~\cite{Aubert:2009aw} and Eq.~\eqref{eq:SpsiK} we get, at $2\sigma$,
\begin{align}
	|{S}_\ell^+| < 0.24 \, .
\end{align}

\section{Isolating parameters of interest}
\label{sec:propose}
Combinations of observables measured by \babar allow us to constrain various parameters of interest. We here use \babar's notation. The combinations that correspond to the CP-odd $C_{\psi K}^e$ and $S_{\psi K}^e$, defined in Eq. (\ref{eq:cgse}), can be isolated via the following combinations:
\beqa
-\frac{1}{2}\left(C_{\ell^+,K_S}^\pm+C_{\ell^+,K_L}^\pm\right)&=&
\frac{1}{2}\left(C_{\ell^-,K_S}^\pm+C_{\ell^-,K_L}^\pm\right)=
\hat C_{\psi K}+\hat\theta^R_{\psi K}\,,\no\\
\frac{\pm1}{2}\left(S_{\ell^+,K_S}^\pm-S_{\ell^+,K_L}^\pm\right)&=&
\frac{\mp1}{2}\left(S_{\ell^-,K_S}^\pm-S_{\ell^-,K_L}^\pm\right)=
\hat S_{\psi K}-\hat G_{\psi K}\hat\theta^I_{\psi K}\,.
\eeqa
Upper bounds on combinations of the wrong-sign T-odd parameters $S_\ell^-$ and $G_\ell^-$, defined in Eq. (\ref{eq:DefLeptons}), and the CPT violating parameter $z$, defined in Eq.~(\ref{eq:defpqz}), can be obtained as follows:
\beqa
\frac{-1}{2}\left(C_{\ell^+,K_S}^\pm+C_{\ell^-,K_S}^\pm\right)&=&
S_{\psi K_S}\left(S_{\ell}^- -z^I\right)\,,\no\\
\frac{-1}{2}\left(C_{\ell^+,K_L}^\pm+C_{\ell^-,K_L}^\pm\right)&=&
S_{\psi K_L}\left(S_{\ell}^- -z^I\right)\,,\no\\
\frac{\pm1}{2}\left(S_{\ell^+,K_S}^\pm+S_{\ell^-,K_S}^\pm\right)&=&
G_{\psi K_S}S_{\psi K_S}\left(G_{\ell}^- -z^R\right)\,,\no\\
\frac{\pm1}{2}\left(S_{\ell^+,K_L}^\pm+S_{\ell^-,K_L}^\pm\right)&=&
G_{\psi K_L}S_{\psi K_L}\left(G_{\ell}^- -z^R\right)\,,
\eeqa

From the results of~\cite{Lees:2012uka} we can get
\beqa
&&\hat S_{\psi K}-\hat G_{\psi K}\hat\theta^I_{\psi K} = 0.69\pm0.04 \, .
\eeqa
and the following bounds can be deduced
\beqa
&&|\hat C_{\psi K}+\hat\theta^R_{\psi K}|<0.07 \, , \no\\
&&|G_{\psi K_{S,L}}S_{\psi K_{S,L}}\left(G_{\ell}^- -z^R\right)|<0.10 \, , \no\\
&&|S_{\psi K_{S,L}}\left(S_{\ell}^- -z^I\right)|<0.06 \, ,
\eeqa
at $2 \sigma$ level.  In case we assume no CPT violation naive combination of the above will lead to
\beqa
|S^-_\ell| < 0.10\,,\;\;\;\;\;\;|G^-_\ell| < 0.21\,,
\eeqa
at $2\sigma$ level.

\section{Experimental asymmetries} \label{sec:rates}
In this Appendix, we provide the full expressions for the asymmetries measured at \babar. For the relation between our notation (\ref{eq:master}) and those of Table I of \cite{Lees:2012uka}, see Eq. (\ref{eq:defdstdct}) and the discussion below it. The asymmetries measured by \babar are the following:
\beqa
\Delta S_T^+&=&\frac{S_{(\psi K_L)_\perp,\ell^-X}}{\kappa_{(\psi K_L)_\perp,\ell^-X}}
-\frac{S_{(\ell^+ X)_\perp,\psi K_S}}{\kappa_{(\ell^+ X)_\perp,\psi K_S}}\,,\ \ \
\Delta C_T^+=\frac{C_{(\psi K_L)_\perp,\ell^-X}}{\kappa_{(\psi K_L)_\perp,\ell^-X}}
-\frac{C_{(\ell^+ X)_\perp,\psi K_S}}{\kappa_{(\ell^+ X)_\perp,\psi K_S}}\,,\nonumber\\
\Delta S_T^-&=&\frac{S_{(\ell^- X)_\perp,\psi K_L}}{\kappa_{(\ell^- X)_\perp,\psi K_L}}
-\frac{S_{(\psi K_S)_\perp,\ell^+X}}{\kappa_{(\psi K_S)_\perp,\ell^+X}}\,,\ \ \
\Delta C_T^-=\frac{C_{(\ell^- X)_\perp,\psi K_L}}{\kappa_{(\ell^- X)_\perp,\psi K_L}}
-\frac{C_{(\psi K_S)_\perp,\ell^+X}}{\kappa_{(\psi K_S)_\perp,\ell^+X}}\,,\nonumber\\
\Delta S_{CP}^+&=&\frac{S_{(\ell^- X)_\perp,\psi K_S}}{\kappa_{(\ell^- X)_\perp,\psi K_S}}
-\frac{S_{(\ell^+ X)_\perp,\psi K_S}}{\kappa_{(\ell^+ X)_\perp,\psi K_S}}\,,\ \ \
\Delta C_{CP}^+=\frac{C_{(\ell^- X)_\perp,\psi K_S}}{\kappa_{(\ell^- X)_\perp,\psi K_S}}
-\frac{C_{(\ell^+ X)_\perp,\psi K_S}}{\kappa_{(\ell^+ X)_\perp,\psi K_S}}\,,\nonumber\\
\Delta S_{CP}^-&=&\frac{S_{(\psi K_S)_\perp,\ell^-X}}{\kappa_{(\psi K_S)_\perp,\ell^-X}}
-\frac{S_{(\psi K_S)_\perp,\ell^+X}}{\kappa_{(\psi K_S)_\perp,\ell^+X}}\,,\ \ \
\Delta C_{CP}^-=\frac{C_{(\psi K_S)_\perp,\ell^-X}}{\kappa_{(\psi K_S)_\perp,\ell^-X}}
-\frac{C_{(\psi K_S)_\perp,\ell^+X}}{\kappa_{(\psi K_S)_\perp,\ell^+X}}\,,\nonumber\\
\Delta S_{CPT}^+&=&\frac{S_{(\psi K_L)_\perp,\ell^+X}}{\kappa_{(\psi K_L)_\perp,\ell^+X}}
-\frac{S_{(\ell^+ X)_\perp,\psi K_S}}{\kappa_{(\ell^+ X)_\perp,\psi K_S}}\,,\ \ \
\Delta C_{CPT}^+=\frac{C_{(\psi K_L)_\perp,\ell^+X}}{\kappa_{(\psi K_L)_\perp,\ell^+X}}
-\frac{C_{(\ell^+ X)_\perp,\psi K_S}}{\kappa_{(\ell^+ X)_\perp,\psi K_S}}\,,\nonumber\\
\Delta S_{CPT}^-&=&\frac{S_{(\ell^+ X)_\perp,\psi K_L}}{\kappa_{(\ell^+ X)_\perp,\psi K_L}}
-\frac{S_{(\psi K_S)_\perp,\ell^+X}}{\kappa_{(\psi K_S)_\perp,\ell^+X}}\,,\ \ \
\Delta C_{CPT}^-=\frac{C_{(\ell^+ X)_\perp,\psi K_L}}{\kappa_{(\ell^+ X)_\perp,\psi K_L}}
-\frac{C_{(\psi K_S)_\perp,\ell^+X}}{\kappa_{(\psi K_S)_\perp,\ell^+X}}\,.
\eeqa
We find the following expressions for these asymmetries:
\beqa
\Delta S_T^+&=&-\Delta S_T^-=-2\left[\hat S_{\psi K}\left(1+\hat G_{\psi K}\left(G_\ell^- -z^R\right)\right)-\hat G_{\psi K}\hat\theta_{\psi K}^I\right]\,,\no\\
\Delta C_T^+&=&\Delta C_T^-=2\left[\hat C_{\psi K}+\hat S_{\psi K}\left( S_\ell^-
-z^I\right)+\hat\theta_{\psi K}^R\right]\,,\no\\
\Delta S_{CP}^+&=&-\Delta S_{CP}^-=-2\left[S_{\psi K_S}-G_{\psi K_S}\theta_{\psi K_S}^I+S_{\psi K_S}G_{\psi K_S}G_\ell^+ -z^I\left(1-\hat S_{\psi K}^2\right)\right]\,,\no\\
\Delta C_{CP}^+&=&\Delta C_{CP}^-=2\left[C_{\psi K_S}+S_{\psi K_S}S_\ell^+ +\theta_{\psi K_S}^R+G_{\psi K_S}z^R\right]\,,\no\\
\Delta S_{CPT}^+&=&-\Delta S_{CPT}^-=-2\left[\Delta S_{\psi K}-z^I\left(1-\hat S_{\psi K}^2\right)+\hat G_{\psi K}\left(\hat S_{\psi K}G_{\ell^+}-\Delta\theta^I_{\psi K}-\hat S_{\psi K}z^R\right)\right]\,,\no\\
\Delta C_{CPT}^+&=&\Delta C_{CPT}^-=2\left[\Delta C_{\psi K}+\Delta\theta^R_{\psi K}+\hat S_{\psi K}\left(S_{\ell^+}-z^I\right)+\hat G_{\psi K}z^R\right]\,.
\eeqa
%
We notice that not only do the T-asymmetries get T-even contributions, as explained in Section~\ref{sec:babar}, but also the CPT asymmetries get CPT-even contributions. All of these effects vanish if there is neither CPT violation in strangeness changing decays nor wrong strangeness decays nor wrong sign decays.

\section{Theoretical asymmetries} \label{sec:ratest}
It is interesting to define ``theoretical" asymmetries where the initial states of the corresponding experimental asymmetries are replaced by the time-conjugate of the final states. Thus, instead of the experimental method, of observing one of two entangled $B$-mesons decaying, thus projecting the other $B$-meson onto the orthogonal state, here we refer to corresponding ``gedanken experiments", that start with inverse decays:
\beqa\label{eq:defsthe}
\Delta S_T^{+(t)}&=&\frac{S_{(\psi K_S)^T,\ell^-X}}{\kappa_{(\psi K_S)^T,\ell^-X}}
-\frac{S_{(\ell^- X)^T,\psi K_S}}{\kappa_{(\ell^- X)^T,\psi K_S}}\,,\ \ \
\Delta C_T^{+(t)}=\frac{C_{(\psi K_S)^T,\ell^-X}}{\kappa_{(\psi K_S)^T,\ell^-X}}
-\frac{C_{(\ell^- X)^T,\psi K_S}}{\kappa_{(\ell^- X)^T,\psi K_S}}\,,\nonumber\\
\Delta S_T^{-(t)}&=&\frac{S_{(\ell^+ X)^T,\psi K_L}}{\kappa_{(\ell^+ X)^T,\psi K_L}}
-\frac{S_{(\psi K_L)^T,\ell^+X}}{\kappa_{(\psi K_L)^T,\ell^+X}}\,,\ \ \
\Delta C_T^{-(t)}=\frac{C_{(\ell^+ X)^T,\psi K_L}}{\kappa_{(\ell^+ X)^T,\psi K_L}}
-\frac{C_{(\psi K_L)^T,\ell^+X}}{\kappa_{(\psi K_L)^T,\ell^+X}}\,,\nonumber\\
\Delta S_{CP}^{+(t)}&=&\frac{S_{(\ell^+ X)^T,\psi K_S}}{\kappa_{(\ell^+ X)^T,\psi K_S}}
-\frac{S_{(\ell^- X)^T,\psi K_S}}{\kappa_{(\ell^- X)^T,\psi K_S}}\,,\ \ \
\Delta C_{CP}^{+(t)}=\frac{C_{(\ell^+ X)^T,\psi K_S}}{\kappa_{(\ell^+ X)^T,\psi K_S}}
-\frac{C_{(\ell^- X)^T,\psi K_S}}{\kappa_{(\ell^- X)^T,\psi K_S}}\,,\nonumber\\
\Delta S_{CP}^{-(t)}&=&\frac{S_{(\psi K_L)^T,\ell^-X}}{\kappa_{(\psi K_L)^T,\ell^-X}}
-\frac{S_{(\psi K_L)^T,\ell^+X}}{\kappa_{(\psi K_L)^T,\ell^+X}}\,,\ \ \
\Delta C_{CP}^{-(t)}=\frac{C_{(\psi K_L)^T,\ell^-X}}{\kappa_{(\psi K_L)^T,\ell^-X}}
-\frac{C_{(\psi K_L)^T,\ell^+X}}{\kappa_{(\psi K_L)^T,\ell^+X}}\,,\nonumber\\
\Delta S_{CPT}^{+(t)}&=&\frac{S_{(\psi K_S)^T,\ell^+X}}{\kappa_{(\psi K_S)^T,\ell^+X}}
-\frac{S_{(\ell^- X)^T,\psi K_S}}{\kappa_{(\ell^- X)^T,\psi K_S}}\,,\ \ \
\Delta C_{CPT}^{+(t)}=\frac{C_{(\psi K_S)^T,\ell^+X}}{\kappa_{(\psi K_S)^T,\ell^+X}}
-\frac{C_{(\ell^- X)^T,\psi K_S}}{\kappa_{(\ell^- X)^T,\psi K_S}}\,,\nonumber\\
\Delta S_{CPT}^{-(t)}&=&\frac{S_{(\ell^- X)^T,\psi K_L}}{\kappa_{(\ell^- X)^T,\psi K_L}}
-\frac{S_{(\psi K_L)^T,\ell^+X}}{\kappa_{(\psi K_L)^T,\ell^+X}}\,,\ \ \
\Delta C_{CPT}^{-(t)}=\frac{C_{(\ell^- X)^T,\psi K_L}}{\kappa_{(\ell^- X)^T,\psi K_L}}
-\frac{C_{(\psi K_L)^T,\ell^+X}}{\kappa_{(\psi K_L)^T,\ell^+X}}\,.
\eeqa
We use the same approximations as in Sec.~\ref{sec:tra}. We find:
\beqa
\Delta S_T^{+(t)}&=&-2S_{\psi K_S}\left[1-G_{\psi K_S}\left(z^R+G_\ell^+\right)\right]\,,\no\\
\Delta S_T^{-(t)}&=&-2S_{\psi K_L}\left[1+G_{\psi K_L}\left(z^R-G_\ell^+\right)\right]\,,\no\\
\Delta C_T^{+(t)}&=&2\left[C_{\psi K_S}-S_{\psi K_S}\left(z^I+S_\ell^+\right)\right]\,,\no\\
\Delta C_T^{-(t)}&=&2\left[C_{\psi K_L}+S_{\psi K_L}\left(z^I-S_\ell^+\right)\right]\,,\no\\
\Delta S_{CP}^{+(t)}&=&-2\left[S_{\psi K_S}-G_{\psi K_S}\theta^I_{\psi K_S}-S_{\psi K_S}G_{\psi K_S}G_\ell^+ -z^I\left(1-S_{\psi K_S}^2\right)\right]\,,\no\\
\Delta S_{CP}^{-(t)}&=&-2\left[S_{\psi K_L}+G_{\psi K_L}\theta^I_{\psi K_L}-S_{\psi K_L}G_{\psi K_L}G_\ell^+ +z^I\left(1-S_{\psi K_L}^2\right)\right]\,,\no\\
\Delta C_{CP}^{+(t)}&=&2\left[C_{\psi K_S}-S_{\psi K_S}S_\ell^++\theta_{\psi K_S}^R+G_{\psi K_S}z^R\right]\,,\no\\
\Delta C_{CP}^{-(t)}&=&2\left[C_{\psi K_L}-S_{\psi K_L}S_\ell^--\theta_{\psi K_L}^R-G_{\psi K_L}z^R\right]\,,\no\\
\Delta S_{CPT}^{+(t)}&=&2\left[z^I\left(1-S_{\psi K_S}^2\right)+G_{\psi K_S}\left(\theta_{\psi K_S}^I+S_{\psi K_S}z^R\right)\right]\,,\no\\
\Delta S_{CPT}^{-(t)}&=&-2\left[z^I\left(1-S_{\psi K_L}^2\right)+G_{\psi K_L}\left(\theta_{\psi K_L}^I+S_{\psi K_L}z^R\right)\right]\,,\no\\
\Delta C_{CPT}^{+(t)}&=&2\left[\theta^R_{\psi K_S}-S_{\psi K_S}z^I+G_{\psi K_S}z^R\right]\,,\no\\
\Delta C_{CPT}^{-(t)}&=&-2\left[\theta^R_{\psi K_L}-S_{\psi K_L}z^I+G_{\psi K_L}z^R\right]\,.
\eeqa
As expected, the theoretical T asymmetries have only T-odd contributions, the theoretical CP asymmetries have only CP-odd contributions, and the theoretical CPT asymmetries have only CPT-odd contributions. Furthermore, in the absence of wrong strangeness decays, wrong sign decays and CPT violation in strangeness  changing decays , the theoretical asymmetries equal the corresponding experimental asymmetries.

\section{EPR entanglement with CPT violation}
\label{sec:EPR}

In this appendix we show that the factorization of the decay amplitudes to $e^{-\Gamma(t_1+t_2)}\times f(t_2-t_1)$  holds in the presence of CPT violation. We follow Ch.~9 of~\cite{Branco:1999fs} with the appropriate modifications for the CPT violating case. The initial pair is in a state
\beq
\left| \Phi^c \right\rangle = \frac{1}{\sqrt{2}}\left[ |B(\vec{k})\rangle\otimes|\overline{B}(-\vec{k})\rangle-
|\overline{B}(\vec{k})\rangle\otimes|B(-\vec{k})\rangle \right] \, ,
\eeq
where the relative (-) sign is a result of the C-parity of the $\Upsilon(4S)$. The $\pm\vec{k}$ are the three momenta of the left and right moving meson in the resonance rest frame.

The decay amplitude of the meson with momenta $\vec{k}$ into final state $f_1$ and of the one with momenta $-\vec{k}$ to final state $f_2$ at times $t_1$ and $t_2$, respectively, is $ (t\equiv t_2-t_1)$
\beq \label{eq:Amp}
\langle f_1,t_1\,; f_2, t_2|T|\Phi^c\rangle
%
=&&\frac{  e^{-(\Gamma/2+im)( t_1+t_2)  }  }{\sqrt{2}} \times \no\\
\Bigg[&&\left(\sqrt{1-z^2}\left(\frac{p}{q}A_{f_1}A_{f_2}-\frac{q}{p}\bar{A}_{f_1}\bar{A}_{f_2} \right)
+z(\bar{A}_{f_2}A_{f_1}+\bar{A}_{f_1}A_{f_2})  \right)g_-(t)  \, ,\no\\
&&+(\bar{A}_{f_2}A_{f_1}-\bar{A}_{f_1}A_{f_2})g_+(t)\Bigg] \, ,
\eeq
where
\beq
&&g_+(t) =  \cos\left(\frac{x\Gamma t}{2}\right)\cosh\left(\frac{y\Gamma t}{2}\right)
+i\sin\left(\frac{x\Gamma t}{2}\right)\sinh\left(\frac{y\Gamma t}{2}\right) \, , \no\\
&&g_-(t) = \cos\left(\frac{x\Gamma t}{2}\right)\sinh\left(\frac{y\Gamma t}{2}\right)+i\sin\left(\frac{x\Gamma t}{2}\right)\cosh\left(\frac{y\Gamma t}{2}\right)\, .
\eeq
By squaring the absolute value of the amplitude in Eq.~\eqref{eq:Amp} we get that it factorizes as $e^{-\Gamma(t_1+t_2)}\times f(t)$.



\end{document}